\documentstyle[12pt]{article}      
                                        
\newcommand{\Od}{{\cal O}}

\newcommand{\qmixv}{(q_0,\vec{q},\tau)}
\newcommand{\qmixw}{(q_0,\omega_q,\tau)}
\newcommand{\tr}{\mbox{tr}}

\newcommand{\im}{\mbox{Im}}

\newcommand{\sgn}{\mbox{sgn}}

\newcommand{\fpi}{f_\pi}
\newcommand{\fpite}{f_\pi^t}
\newcommand{\fpisp}{f_\pi^s}
\newcommand{\ft}{f(t)}
\newcommand{\fpit}{f_\pi(t)}
\newcommand{\gpit}{g_\pi (t)}

\newcommand{\fpitsq}{f_\pi^2 (t)}
\newcommand{\ftsq}{f^2(t)}
\newcommand{\fzerosq}{f^2(0)}
\newcommand{\fdot}{\dot f(t)}
\newcommand{\fddot}{\ddot f(t)}

\newcommand{\tti}{\tilde t}

\newcommand{\intc}{\int_C dt \int d^3 \vec{x}}
\newcommand{\intcc}{\int_C d^4x}
\newcommand{\vxt}{(\vec{x},t)}

\newcommand{\be}{\begin{equation}}
\newcommand{\ee}{\end{equation}}
\newcommand{\ba}{\begin{eqnarray}}
\newcommand{\ea}{\end{eqnarray}}

\newcommand{\NP}[1]{{\it Nucl.\ Phys.\ }{\bf #1}}
\newcommand{\ZP}[1]{{\em Z.\ Phys.\ }{\bf #1}}

\newcommand{\PL}[1]{{\em Phys.\ Lett.\ }{\bf #1}}

\newcommand{\AN}[1]{{\em Ann. Phys. } (N.Y.) {\bf #1}}

\newcommand{\PRep}[1]{{\em Phys.\ Rep.\ }{\bf #1}}
\newcommand{\PR}[1]{{\em Phys.\ Rev.\ }{\bf #1}}

\newcommand{\IJmp}[1]{{\em Int.\ J.\ Mod.\ Phys.\ }{\bf #1}}

\def\IN{\relax{\rm I\kern-.18em N}}
\def\IR{\relax{\rm I\kern-.18em R}}
\def\ID{\relax{\rm I\kern-.18em D}}

\newcommand{\dnote}[1]{}
\begin{document}
\input{epsf}

\begin{flushright}
FT/UCM/2-98

IMPERIAL/TP/98-99/14

\end{flushright}

\vspace{.75cm}

\begin{center} 
{\large\bf NONEQUILIBRIUM 
CHIRAL PERTURBATION THEORY AND PION DECAY FUNCTIONS}\\
\vskip 1.2cm
{\large 
A. G\'{o}mez Nicola$^a
$\footnote{E-mail:{\tt  gomez@eucmax.sim.ucm.es}}
and  V.Gal\'an-Gonz\'alez$^b$\footnote{E-mail: 
{\tt v.galan-gonz@ic.ac.uk}}
}\\
\vskip 5pt 
{\it a}) 
Departamento de F\'{\i}sica 
Te\'orica,  Universidad Complutense, 28040, Madrid, Spain
\vskip 3pt
{\it b}) Theoretical Physics, 
Blackett Laboratory, Imperial College, Prince Consort Road, London,
SW7 2BZ, United Kingdom\\
\end{center}

\today

\begin{abstract}

We extend   chiral perturbation theory  to study 
 a meson  gas out of thermal equilibrium. Assuming that  the system is  
initially in  equilibrium  at  $T_i<T_c$ and working  
within the   Schwinger-Keldysh contour technique, we  define consistently
 the time-dependent  temporal and spatial pion decay functions, 
 the counterparts of  the pion decay constants, 
 and  calculate  them   to  next to leading order. The 
  link with  curved  space-time  QFT allows 
 to establish nonequilibrium renormalisation.  
  The short-time behaviour and the  applicability of our model to a  
heavy-ion  collision plasma  are also discussed in this work.

\end{abstract}

\vspace{1cm}

\begin{flushleft}
  {\bf PACS numbers:} 12.39.Fe, 11.10.Wx, 05.70.Ln, 12.38.Mh
  
\end{flushleft}
\newpage
\section{Introduction}
\label{intro}
The chiral phase  transition plays a fundamental role in the description of   
 the plasma formed after a  relativistic 
 heavy-ion collision (RHIC), where it is imperative to use meson effective 
 models to describe QCD.  Two of the  most successful approaches
 are the $O(4)$ linear sigma model (LSM), valid only for  $N_f=2$ light 
 flavours, and Chiral Perturbation Theory (ChPT), based on  derivative 
 expansions  compatible with the QCD symmetries, and whose
  lowest order action is the nonlinear sigma model (NLSM)
  \cite{wegale}. In ChPT, 
 the  perturbative parameter is  $p/\Lambda_\chi$,  
 with $p$  a  meson energy 
 (like masses, external momenta or temperature) and 
 $\Lambda_{\chi}\simeq$ 1 GeV.  
  Every meson loop is  $\Od(p^2/\Lambda^2_\chi)$  and all the 
 infinities 
 coming from them can  be absorbed in the coefficients of 
 higher order lagrangians \cite{wegale,dogolope97}.

 In thermal equilibrium at finite temperature $T$,  
the chiral symmetry is believed to be restored at   
$T_c\simeq$ 150-200 MeV  \cite{wi92}. In fact, near $T_c$, the 
mean-field LSM is  well known to undergo a second-order phase transition.
 The NLSM is equally valid for 
 reproducing the phase transition, provided one works 
  in the large-$N$  limit  \cite{boka96}. Strictly
 within ChPT, the low-temperature meson gas has been studied 
 \cite{gale87,gele89} \dnote{paper JR?} 
on expansion in $T^2/\Lambda_{\chi}^2$, predicting 
 the  correct behaviour of the observables 
as $T$ approaches $T_c$.

The equilibrium assumption is not realistic if one is interested 
 in the dynamics of the expanding plasma formed after a RHIC, where 
several  nonequilibrium effects could be important.  
One of them is the 
 formation of disoriented chiral condensates (DCC), 
  regions in which the chiral field is correlated and has nonzero 
 components in the pion direction  \cite{an89}.
 As the plasma expands, long-wavelength pion modes 
 ---propagating as if they had  an effective negative 
mass squared---
can develop  instabilities  growing fast as the  field relaxes to the ground 
 state, an observable consequence being   coherent pion emission 
\cite{rawi93}.   This issue has been 
 extensively studied in the literature, mostly
  within  the LSM   assuming initial  thermal equilibrium 
at $T_i>T_c$,  either  
 encoding  the cooling mechanism in the time dependence of the
 lagrangian parameters [8-11] or describing  the plasma expansion in 
  proper  time and rapidity  \cite{cooper95}. This phenomenon has also
 been studied using   
Gross-Neveu models \cite{bard96}.
  Another  important nonequilibrium  observable  
is  the photon and dilepton production \cite{baier97}, to which the 
 anomalous meson sector could  significantly contribute
\cite{bodeho97}.

In this work we will construct an effective ChPT-based model to describe 
a meson 
 gas out of thermal equilibrium,  as an alternative to the  LSM 
 approach.    Our only degrees of freedom will 
  be then the  Nambu-Goldstone bosons (NGB) 
 and we will consider  the most general 
 low-energy lagrangian  compatible with the QCD  symmetries.
 We will restrict here to  $N_f=2$ (where the NGB are just the pions)   
and to the chiral limit (massless quarks), 
which is   the simplest approximation 
  allowing  to build the model in terms of exact chiral symmetry. 
One of  the novelties of our approach is to exploit the
 analogy between ChPT
 and the physical regime where the system is not far from equilibrium and 
 then a derivative expansion is consistent. 
\section{The NLSM and ChPT out of equilibrium}
\label{model}

\begin{figure}\epsfxsize=15cm
\vspace*{-8cm}
\centerline{\epsfbox{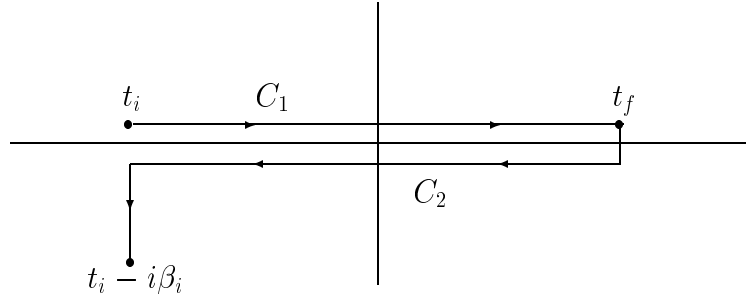}}
\vspace*{-9cm}
\vskip 4mm
\caption {\label{fig1} The contour $C$ in complex time $t$. The 
 lines $C_1$ and $C_2$ run  between $t_i+i\epsilon$ and 
 $t_f+i\epsilon$ and  $t_f-i\epsilon$ and $t_i-i\epsilon$ 
respectively, 
 with  $\epsilon\rightarrow 0^+$.}
\end{figure}
 We will take  the system   in thermal equilibrium for $t\leq0$ at a 
 temperature $T_i<T_c$ and  for $t>0$ we let the  lagrangian parameters 
 be  time-dependent. We are also   assuming that the 
 system is homogeneous 
 and isotropic.  The generating functional 
  of the theory can then be formulated in the  path integral formalism, 
  by letting the  time 
 integrals run over the Schwinger-Keldysh 
contour  $C$ displayed in Fig.\ref{fig1} [15-18]. 
\dnote{check} 
 We will eventually let  $t_i\rightarrow -\infty$ and 
 $t_f\rightarrow +\infty$, although we will show  that 
our results are independent of $t_i$ and $t_f$. 
  We remark that, even in that limit, 
  the imaginary-time leg of $C$  has to be kept, since it 
encodes the KMS equilibrium boundary conditions 
\cite{sewe85,boleesi93,lebe96}. With these 
  assumptions, our low-energy model will be the 
 following nonequilibrium NLSM 
\begin{equation}
S[U]=\intcc \ \frac{f^2(t)}{4} \ \tr \ \partial_\mu U^\dagger\vxt
\partial^\mu U\vxt
\label{nlsmne}
\end{equation}
where $\intcc\equiv\intc$, $U\vxt\in SU(2)$ is 
the NGB field, satisfying 
 $U (\vec{x},t_i+i\beta_i)=U(\vec{x},t_i)$ with $\beta_i=T_i^{-1}$, and 
$f(t)$ is a real function which in equilibrium and to the 
 lowest order (see section \ref{fpi})
 would be $f=f_\pi\simeq 93$ MeV (the pion decay constant) i.e, 
   \mbox{$f(t\leq 0)=f$}. Note that 
 $\ft$ cannot be  analytic at $t=0$ and, in particular, it 
could be discontinuous, like the meson mass in  
quenched LSM approaches [8-11].
 This is a consequence of the nature of our approach, since the system is 
 driven off equilibrium {\it instantaneously}. An alternative, which we 
will not attempt here, is to choose $\ft$
 analytic $\forall t$, having equilibrium only at $t=t_i$ 
 \cite{agnwip}\dnote{adiabatically?}. Thus,  
the temporal  evolution  of our results  will   start 
at $t=0^+$,  an infinitesimally   small response  time. As we will see below, 
 our approach is consistent because the discontinuities at $t=0$ appear 
 to  NLO. 

 We will parametrise the field $U$  as
\ba
U\vxt=\frac{1}{f(t)}\left\{\left[f^2(t)-\pi^2\vxt \right]^{1/2} I
+i\tau_a\pi^a\vxt\right\}\quad ; \quad a=1,2,3,
\label{upar}
\ea
where $\pi^2=\pi^a\pi_a$,  $\pi^a$ the pion fields satisfying 
 $\pi^a (t_i+i\beta_i)=\pi^a (t_i)$  and   
$I$ and $\tau_a$ are the identity and Pauli matrices. 
 Note that with the choice 
 (\ref{upar})  we recover the 
 canonical kinetic term in the action after expanding  $U$ in powers of $\pi$.
 Other choices  amount to a 
 time-dependent normalisation of the pion fields and  should not have 
 any effect on the physics  (see section \ref{fpi}). 
 For instance, if we  redefine 
 $\tilde\pi^a=\pi^a f(0)/\ft$, the action  
  for the $\tilde\pi^a\vxt$ fields  is 
  just the  equilibrium  NLSM multiplied by the  time-dependent scale 
factor $\ftsq/\fzerosq$ (see below).


 Our action (\ref{nlsmne}) is manifestly chiral invariant 
(\mbox{$U(x)\rightarrow LU(x)R^\dagger$}). 
 Notice that we work in the chiral limit and hence 
 there are no explicit symmetry-breaking pion mass terms in the action. 
  The conserved axial and vector  currents for the chiral symmetry can be 
 derived  by applying the standard procedure  \cite{wegale,dogolope97}, so
 that  the axial current reads 
\begin{eqnarray}
A_\mu^a\vxt&=&i\frac{\ftsq}{4}\tr\left[\tau^a\left(
U^\dagger\partial_\mu U-U\partial_\mu U^\dagger\right)\right]
\label{currents}
\end{eqnarray}

Let us now  discuss how to establish a consistent nonequilibrium ChPT. 
The new ingredient we need  is the temporal variation of $\ft$. We will then
 consider 
\begin{equation}
\frac{\fdot}{f^2 (t)}\simeq \Od\left(\frac{p}{\Lambda_\chi}\right), \qquad 
\frac{\fddot}{f^3(t)}, \frac{[\fdot]^2}{f^4(t)}
\simeq\Od \left(\frac{p^2}{\Lambda_\chi^2}\right), 
\label{chipoco}  
\end{equation}
 and so on, 
  the rest of the chiral power counting being the same as in equilibrium. 
  Therefore, in our approach we treat  the deviations of 
 the system from equilibrium perturbatively, following  the ChPT guidelines. 
 Thus, 
we will  expand our action 
(\ref{nlsmne}) 
 to the relevant order in pion fields and take into account 
 all the contributing Feynman diagrams. 
The loop  divergences should be such that they can be absorbed 
 in the coefficients of  higher order lagrangians, 
which in general will require 
the introduction of new time-dependent 
  counterterms (see below). Notice also 
 that according to  (\ref{chipoco}), we can always 
 describe the short-time nonequilibrium regime, just by 
 expanding $\ft$ around $t=0^+$. In fact, 
for times $t\leq \fpi^{-1}$, that  is equivalent to a chiral expansion,
 since then $\dot f (0^+) t/f(0^+)=\Od (p/\Lambda_\chi)$ 
and so on. 
 Nonetheless,  we  stress that the 
  conditions (\ref{chipoco}) do not imply working at short-times,
 but just to remain  close enough to equilibrium.


 To leading order in $\pi$ fields, 
the action (\ref{nlsmne}), after using (\ref{upar}),  reads
\begin{eqnarray}
S_0[\pi]&=&- \frac{1}{2}\intcc\pi^a\vxt\left[  \Box + 
m^2 (t)\right] 
\pi^a\vxt \label{actlo}  \quad \mbox{with} \quad 
m^2(t)=-\frac{\fddot}{\ft},
\end{eqnarray}
where we have partial integrated in  $C$. 
 Thus,  the leading order nonequilibrium effect of our model can 
 be written as   a time-dependent pion mass term, which,  
as  commented before, is   a  common  
   feature of nonequilibrium models  [9-12].
\dnote{check} 
Notice that $m^2(t)$ can be 
 negative, so that our model accommodates  unstable pion 
modes, whose importance we have discussed before. 
 Note also  that this mass term does {\it not} 
break the chiral symmetry, i.e,    
 the axial current is classically   conserved. Indeed, 
to leading order  we have, from  (\ref{currents}), 
 \be
\left[A_\mu^a\vxt\right]^{LO}=
-\ft\partial_\mu\pi^a\vxt+\delta_{\mu 0}\fdot\pi^a\vxt 
\quad ,
\label{axcurrlo}
\ee
which satisfies $\partial^\mu A_\mu=0$ using 
 $\left[\Box + m^2 (t)\right] \pi^a=0$, the   
 equations of motion to the same order.   
 Had we included the  
 pion mass term $m_\pi$ ---explicitly breaking the symmetry---
 the instabilities threshold, to leading order, 
  would have been   $m^2(t)<-m_\pi^2$ instead.

 It is very interesting to  rephrase our model 
 as a NLSM in a curved space-time background
 $g_{\mu\nu}$, which reads   \cite{bida82,dole91},  
\ba
S_g[U]= \frac{f^2(0)}{4}\intcc \left(\sqrt{- g}\right)  g^{\mu\nu} 
 \ \tr \ \partial_\mu U^\dagger (x)
\partial_\nu U (x)+\xi S_R[U,R]
\label{nlsmcurv}
\ea
plus $U$ independent terms, where $g=\det g$ and  the last term 
 accounts for  possible  
couplings between the  pion fields and the 
 scalar curvature $R (x)$ (like  $R(x)\phi^2$  for a free scalar field 
$\phi$  \cite{bida82}). 
 Now, notice that  our nonequilibrium model  (\ref{nlsmne})
is obtained by writing 
 $U(x)$ in the $\tilde\pi$ parametrisation discussed before (i.e, with 
 $f(t)$ replaced by $f(0)$ in (\ref{upar})), 
   choosing $\xi=0$ (minimal coupling)  and  a 
 spatially flat Robertson-Walker (RW)  space-time in conformal time, 
 whose line element is $ds^2=a^2(\eta)[d\eta^2-d\vec{x}^2]$,  
with the scale factor 
 $a(\eta)=f(\eta)/f(0)$. 
 Our effective theory  is then not only suitable 
 for a RHIC environment, but also in a cosmological 
 framework. 
 Notice also that if we take   $\xi\neq 0$,  the lowest order  $S_R$ 
term we can construct has the form of an effective mass term breaking 
 explicitly the chiral symmetry. 
In fact, it is not difficult to see that
 we could cancel  the $m^2(t)$ term in (\ref{actlo})  by  choosing 
 $\xi=1/6$, which  is  the value
 rendering the theory scale invariant 
\cite{bida82}. This is just a consequence of 
 the lagrangian chiral and conformal 
 symmetries   being  incompatible in a curved background 
\cite{dole91} or,
 equivalently, at  nonequilibrium.  In other words, 
 for $\xi=0$ ---which is our choice, since we want to preserve chiral 
 symmetry, as in \cite{dole91}---
 we may interpret the $m^2(t)$ term, in the chiral limit,   
as the minimal coupling with the background  yielding  chiral invariance.

The above equivalence turns out to be very useful  to renormalise our model,
 consistently with ChPT. In fact,  all the one-loop divergences arising 
 from  (\ref{nlsmcurv}) can be absorbed 
 in the coefficients of the $\Od (p^4)$ action $S_4$, which consists of  
 the  Minkowski terms with indices raised and lowered with 
 $g_{\mu\nu}$ plus  new chiral-invariant 
couplings of pion fields  with the  curvature \cite{dole91}. 
In the chiral limit,  those new terms read 
\ba
S_4^R\left[\pi\right]&=&\intcc \left(\sqrt{- g}\right)\left[
L_{11}R(x)g^{\mu\nu}+L_{12}R^{\mu\nu}\right]\tr 
\partial_\mu U^\dagger (x)
\partial_\nu U (x)
\nonumber\\&=&
 - \frac{1}{2}\intcc\pi^a\left[f_1 (t)\partial_t^2-f_2 (t)\nabla^2
 + m_1^2 (t)\right]\pi^a + \Od (\pi^4)
\label{s4}
\ea
where $R_{\mu\nu}$ is the Ricci tensor,  $L_{11}$ and $L_{12}$ are two 
 new  low-energy constants  \cite{dole91} and  we 
have given the two-pion contributions in the parametrisation 
(\ref{upar}), after partial integration, with our RW metric, where 
\ba
f_1(t)&=&12\left[\left(2L_{11}+L_{12}\right)\frac{\fddot}{f^3(t)}
-L_{12}\frac{[\fdot]^2}{f^4 (t)}\right]
\nonumber\\
f_2(t)&=&4\left[\left(6L_{11}+L_{12}\right)\frac{\fddot}{f^3(t)}
+L_{12}\frac{[\fdot]^2}{f^4 (t)}\right]
\nonumber\\
m_1^2 (t)&=&-\left[\frac{f_1(t)\fddot+\dot f_1 (t)\fdot}{\ft}+
\frac{1}{2}\ddot f_1 (t)\right].
\label{f12}
\ea
for $t>0$ and $f_i(t\leq 0)=0$.  
The above terms are the only ones in $S_4$ containing two pions and 
they will renormalise purely nonequilibrium infinities 
---which are time-dependent and vanish for 
 $t\leq0$---. It is important to bear in mind that to cancel the one-loop
 new divergences only $L_{11}$ needs to be renormalised, whereas 
 $L_{12}=L_{12}^r$ \cite{dole91}. 
    We will come back to this point below.

 Next, we will  concentrate on 
the Green functions  time-ordered  along  $C$  
\cite{sewe85,boleesi93}. Unless otherwise
 stated, we will be using the parametrisation (\ref{upar}) in the remaining  
 of this work.  The two-point 
 function defines the pion propagator 
$G^{ab} (x,y)=-i < T_C \pi^a (x) \pi^a (y) >$, which  
  to leading order 
$G_0^{ab} (x,y)=\delta^{ab}G_0(x,y)$, by isospin invariance, and 
\begin{equation}
\left\{\Box_x + m^2(x^0)\right\} G_0 (x,y)=-\delta_C (x^0-y^0)
\delta^{(3)} (\vec{x}-\vec{y})
\label{loprop}
\end{equation}
with   KMS equilibrium  conditions
$G^>_0 (\vec{x},t_i-i\beta_i;y)=G^<_0 (\vec{x},t_i;y)$, 
 the advanced and retarded propagators being defined 
as customarily along  $C$.
Notice 
 that $G(x,x')=G(t,t',\vec{x}-\vec{x'})$ due to
 the  nonequilibrium lack of time translation invariance. 
Therefore, we will  define, as 
 customarily,  the ``fast'' temporal variable $t-t'$ and the ``slow'' one 
 $\tau\equiv (t+t')/2$,   so that 
$F(q_0,\omega_q,\tau)$ and $F(\omega_q,t,t')$, with 
$\omega_q^2= |\vec{q}|^2$,  will 
denote, respectively, the 
 fast and mixed (in which only the spatial coordinates are transformed) 
 Fourier transforms of $F(x,x')$. Note that 
 $F(q_0,\omega_q,\tau)$  depends 
 separately  on  $q_0$ and $\omega_q$ because of the thermal loss of 
 Lorentz covariance  and has the extra nonequilibrium $\tau$-dependence.
\dnote{This last sentence could be cut} 
 Then,  in the mixed representation, (\ref{loprop}) becomes
\be
\left[\frac{d^2}{dt^2}+\omega_q^2+m^2(t)\right] G_0 (\omega_q,t,t')
=-\delta_C (t-t')
\label{lopropmix}
\ee

The general solution of (\ref{lopropmix}) is only known  explicitly  for 
 some particular choices of $m^2(t)$ \cite{bida82,sewe85,boleesi93}. 
Formally, we can 
 write it  as a 
 Schwinger-Dyson  equation  as 
\ba
G_0(\omega_q,t,t')=G_0^{eq} (\omega_q, t-t')
+\int_C dz m^2(z) G_0^{eq} (\omega_q, t-z) G_0(\omega_q,z,t') 
\label{sdg0eq}
\ea
with $G_0^{eq} (\omega_q, t-t')$ the equilibrium 
solution of (\ref{lopropmix}), i.e,  with 
 $m^2(t)=0$. \dnote{Remove the SD?}

Another object of interest for our purposes is the Lehman spectral 
function $\rho (x,y)=G^> (x,y)-G^< (x,y)$  \cite{chin85}, 
which  in equilibrium  to leading order is    
$\rho_0^{eq} (q)=-2\pi i \sgn (q_0) \delta (q^2)$
 \cite{lebe96}. Note that,  by construction,  
$G^>(x,y)=G^<(y,x)$, so that 
$\rho(x,y)=-\rho(y,x)$ and  
$\rho (q_0,\omega_q,\tau)=-\rho(-q_0,\omega_q,\tau)$.
The normalisation of $\rho_0$  is 
\begin{equation}
\frac{1}{2\pi i}\int_{-\infty}^{+\infty} q_0 \rho_0(q_0,\omega_q,\tau)=
 \left.\frac{d \rho_0 (\omega_q,t,t')}{dt}\right\vert_{t=t'}=
 -1,
\label{rho0norm}
\end{equation}
which can be readily checked by  using  
(\ref{sdg0eq}) and  $\rho_0 (\omega_q,t,t)=0$.
\dnote{Another property of $\rho$ cut, which I need only if I introduce
 $h_\pi$ later} 
\section{Next to leading order propagator}
\label{nloprop}
\begin{figure}\epsfxsize=15cm
\vspace*{-8cm}
\centerline{\epsfbox{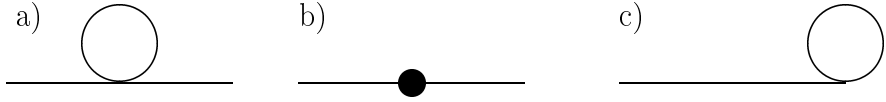}}
\vspace*{-12cm}
\vskip 4mm
\caption {\label{fig2} Diagrams contributing to the 
 NLO pion propagator (a,b) and axial-axial correlator (c). The black dot in 
 b) represents the interaction coming from $S_4^R$ in (\ref{s4})}
\end{figure}

We will now  obtain the NLO correction to the propagator.
 For that purpose, we need the action in 
 (\ref{nlsmne}) up to four-pion terms: 
\begin{eqnarray}
S[\pi]&=&S_0[\pi]+\frac{1}{2}\intcc
\left\{\frac{1}{\ftsq}
\left[ \partial_\mu\pi^a\partial^\mu\pi^b\pi_a\pi_b
\right.\right.\nonumber\\&+&\left.\left.
\frac{1}{2}(\pi^2)^2\left(\frac{\fddot}{\ft}
-\frac{\dot f^2(t)}{\ftsq}\right)
\right] +\Od (\pi^6)\right\}
\label{nlsmnenlo}
\end{eqnarray}
plus the two-pion terms in (\ref{s4}). The two diagrams contributing
 are, respectively, a) and b) in Fig.\ref{fig2}. 

 Let us  concentrate on $G_{11}$ (i.e, $t,t'\in C_1$ in Fig.\ref{fig1})  
 and  $t$ and $t'$ positive, 
which is  the relevant case for our purposes, as commented above. 
We will use dimensional regularisation (DR), so that
 evaluating the above diagrams, using (\ref{loprop}) with 
$\delta^{(d)}(0)=0$, and  after some algebra, we obtain 
 in the  mixed representation to NLO
 \begin{eqnarray}
G_{11}^> (t,t')&=&G_{0,11}^> (t,t')
\left(1-\frac{1}{2}\left[f_1(t)+f_1(t')\right]\right)
+\frac{i}{2\omega_q}
\coth\left[\frac{\beta_i\omega_q}{2}\right] \frac{T_i^2}{12f^2}
\cos\left[\omega_q (t+t')\right]\nonumber\\
&+&i\left\{\int_{0}^t d\tti \left[\Delta_1(\tti,\omega_q)G_0^> (\tti,t)
G_0^>(\tti,t')
+\Delta_2(\tti)\dot G_0^> (\tti,t) \dot G_0^>(\tti,t')\right]\right.
\nonumber\\ 
&-&\int_{0}^{t'} d\tti \left[\Delta_1(\tti,\omega_q)G_0^< (\tti,t)
G_0^<(\tti,t') 
+\Delta_2(\tti)\dot G_0^< (\tti,t) \dot G_0^<(\tti,t')\right]
\nonumber\\
&-&\left.\int_{t'}^t d\tti \left[\Delta_1(\tti,\omega_q)G_0^< (\tti,t)
G_0^>(\tti,t')
+\Delta_2(\tti)\dot G_0^< (\tti,t) \dot G_0^>(\tti,t')\right]\right\}
\label{nloposit}
\end{eqnarray}
and $G_{11}^< (t,t')=G_{11}^> (t',t)$, where 
we have suppressed for simplicity the $\omega_q$ dependence of 
 the propagators,  the dot denotes $d/d\tti$,  
\ba 
\Delta_1 (\tti,\omega_q)&=&\frac{1}{f^2(\tti)}\left[\left(
6\frac{\ddot f(\tti)}{f(\tti)}-
5\left(\frac{\dot f(\tti)}{f(\tti)}\right)^2
-\omega_q^2\right) G_0(\tti)
-2\ddot G_0(\tti)+4\frac{\dot f(\tti)}{f(\tti)}\dot G_0(\tti)
\right]\quad \nonumber\\
&+&i\omega_q^2\left[f_2(\tti)-f_1(\tti)\right]-i\left[\frac{\dot f_1 (\tti)
\dot f(\tti)}{f(\tti)}+\frac{1}{2}\ddot f_1 (\tti)\right]
,\label{del1}\\
\Delta_2(\tti)&=&\frac{G_0(\tti)}{f^2(\tti)}\quad ,
\label{del2}
\ea
 and $G_0(z^0)\equiv G_0(z,z)$ is the equal-time correlation function. 
\dnote{The integral has been cut} 
 We observe that (\ref{nloposit}) is $t_i$ 
and $t_f$ independent, which is a good consistency check. 
 Notice also that 
 by replacing the equilibrium propagators in 
 (\ref{nloposit}),  we recover
\begin{equation}
G_{11}^{eq>} (t-t')=G_{0,11}^{eq>} (t-t')
\left(1-\frac{T^2}{12f^2}\right)
\label{nlogeq}
\end{equation}
which  agrees with   \cite{boka96} (note that we have derived it for 
the contour $C$, including  both imaginary-time and real-time 
 thermal field theory)  and is finite in the 
 chiral limit, where 
there  is no tadpole  renormalisation in DR \cite{dogolope97}. 
  However, out of 
 equilibrium, the NLO propagator is in general  divergent, 
even in the  chiral limit, and the infinities have to  be  
 absorbed in the two-pion  counterterms in (\ref{s4}).   
\section{The nonequilibrium pion decay functions}
\label{fpi}
In a thermal bath, the 
  concepts of LSZ and asymptotic states are subtle, and so is then  the 
 extension of low-energy theorems like PCAC. Thus,  
 pion decay constants are  more 
 conveniently defined through the  thermal axial-axial correlator 
$A_{\mu\nu}^{ab} (x,y)=<T_C A_\mu^a (x) A_\mu^b (y) >$.
 At $T\neq 0$  the  loss of Lorentz covariance in the tensorial structure 
 of $A_{\mu\nu}$  implies that 
one can define  two  independent and complex $\fpisp$ (spatial) 
 and $\fpite$ (temporal),  their real and imaginary parts being  
 related respectively  with  the pion  velocity and damping rate  
 in the thermal bath  \cite{pity96}. 
Nevertheless,  to one-loop in the chiral limit one has 
  \cite{gale87,boka96,pity96}
\begin{equation}
\left[\fpi^s (T)\right]^2=\left[\fpi^t (T)\right]^2
=f^2\left(1-\frac{T^2}{T_c^2}\right),
\label{fpieq}
\end{equation}
with $T_c= \sqrt{6} 
\fpi\simeq$ 228 MeV. Despite it  being just the lowest order  
in the low temperature expansion, (\ref{fpieq}) 
  predicts the right behaviour and a 
reasonable estimate for the critical temperature,  although, strictly 
 speaking,  
$\fpi (T)$ is not the  order parameter   \cite{boka96}.
\dnote{This could be cut} To higher 
 orders, $\fpisp\neq\fpite$ and $\im \fpi^{s,t}\neq 0$  
 \cite{pity96}.



Let us then  analyze $A_{\mu\nu}^{ab}$ in our 
 nonequilibrium model.
 The relevant quantity, as far as 
 $\fpi$ is concerned, is the  spectral function 
$\rho_{\mu\nu}=A^>_{\mu\nu}-A^<_{\mu\nu}$, with 
$A^{ab}_{\mu\nu}=\delta^{ab}A_{\mu\nu}$.  We readily realise that 
$\rho_{\mu\nu} \qmixv=-\rho_{\nu\mu} (-q_0,-\vec{q},\tau)$. Then, 
 from rotational symmetry,  
\be
\rho_{ij} \qmixv = q_i q_j \rho_L \qmixw + \delta_{ij} \rho_d \qmixw
\label{rhoij}
\ee
with $\rho_{L,d}(q_0)=-\rho_{L,d}(-q_0)$ 
\footnote{Our $\rho_L$ and $\rho_d$ correspond in 
in the notation of \cite{boka96},  to 
 $\sgn(q^0)\rho_A^L q_0^2/\omega_q^2 q^2$ and 
$\sgn(q^0)\rho_A^T$ respectively.} and 
$\rho_{j0} \qmixv = q_j \rho_S \qmixw$. 
Therefore, $\rho_{\mu\nu}$ is characterized, in principle by the four 
  functions $\rho_L$, $\rho_d$, $\rho_S$ and $\rho_{00}$. 
 However, they are related through the $A_\mu$ conservation 
Ward Identity (WI)   $\partial^x_\mu \rho^{\mu\nu} (x,y)=\partial^y_\nu 
\rho^{\mu\nu} (x,y)=0$, which also holds  in our model.  Thus, we get
\ba
q^0 \rho_{00}\qmixw -\omega_q^2 \rho_S\qmixw  
-\frac{i}{2}\dot\rho_{00}\qmixw &=&0
\nonumber\\
q^0 \rho_S\qmixw  - \omega_q^2 \rho_L\qmixw 
+\frac{i}{2}\dot\rho_S\qmixw + \rho_d\qmixw &=&0 ,
\label{wirho}
\ea
where the dot denotes $\partial/\partial\tau$. Thus, only two components 
 of $\rho_{\mu\nu}$ 
are independent, as in equilibrium \cite{boka96}, where there 
 are no time derivatives in the above equation.

At $T=0$ one 
 has $\rho_L=2\pi \fpi^2\sgn(q^0)\delta(q^2)$, since there exist NGB states. 
 That is not the case  at $T\neq 0$,  where  the 
pion dispersion relation
 is not in general a $\delta$-function  \cite{boka96}. In fact, 
to define 
 properly $\fpi (T)$ requires taking  the $\omega_q\rightarrow 0^+$ limit, 
 in which  a  zero-energy  excitation still exists  \cite{boka96}, although to
 NLO there is no need to take that limit.   
 Extending  these ideas to nonequilibrium, we will  define the   
 time-dependent   pion decay functions (PDF)  as 
\ba
\left[\fpisp (t)\right]^2&=&\frac{1}{2\pi}\lim_{\omega_q\rightarrow 0^+}
\int_{-\infty}^{\infty} dq_0 q_0 \rho_L (q_0,\omega_q,t)
=\lim_{\omega_q\rightarrow 0^+} i\left.\frac{d}{dt}\rho_L (\omega_q,t,t')
\right\vert_{t=t'}
\label{fpisp}\\
\fpisp (t) \fpite (t) &=&\frac{1}{2\pi}\lim_{\omega_q\rightarrow 0^+}
\int_{-\infty}^{\infty} dq_0  \rho_S (q_0,\omega_q,t)
=\lim_{\omega_q\rightarrow 0^+} \rho_S (\omega_q,t,t)
\label{fpits}\\
\fpisp (t) \gpit &=&-\frac{i}{2\pi} \lim_{\omega_q\rightarrow 0^+} 
 \int_{-\infty}^{\infty} dq_0 q_0  \rho_S (q_0,\omega_q,t)
=\frac{1}{2}\lim_{\omega_q\rightarrow 0^+}
\left.\frac{d}{dt}\rho_S (\omega_q,t,t')
\right\vert_{t=t'}
\label{fpisgpi}
\ea

The functions $\fpisp (t)$ and $\fpite (t)$  are 
the nonequilibrium counterparts  of the spatial and temporal
 pion decay constants respectively, 
whereas $\gpit$ vanishes in equilibrium. 
However, 
the above  PDF  are related through the WI. 
Integrating in $q_0$ in   (\ref{wirho}), we get 
\ba
\fpisp (t) \gpit&=&\frac{1}{2}\frac{d}{dt} 
\left [\fpite (t) \fpisp (t)\right],
\label{wifpi}
\ea
 so that  only two  PDF are independent,  
 as in equilibrium \cite{pity96}. 
    Let us now 
 check the consistency of our    definitions to leading order.  
 From (\ref{axcurrlo}),  
$\rho_L^{LO} (\omega_q,t,t')=if(t) f(t') \rho_0 (\omega_q,t,t')$ 
 and $\rho_d=0$, so that,  using (\ref{rho0norm}) yields  
$\fpisp (t)^2_{LO}=f^2 (t)$, 
i.e, the PDF coincides with $f(t)$ to leading order, as 
 it should be. Similarly, we find 
$\fpite (t)_{LO}=\fpisp (t)_{LO}= \ft$ and $\gpit_{LO}=\dot\ft$, so that our 
 definitions are consistent to leading order. 



To NLO, we need the axial current up to $\Od (\pi^3)$. From (\ref{currents}),
 \ba
A_\mu^ a \vxt =  \left[A_\mu^a\vxt\right]^{LO}
-\frac{1}{2\ft}\left(\pi^a\partial_\mu \pi^2
-\pi^2\partial_\mu \pi^a
 - \delta_{\mu 0} \frac{\fdot}{\ft}\pi^2 \pi^a \right)+
 \Od (\pi^5)
\label{axcurrnlo}
\ea
 with $A_\mu^{LO}$ in (\ref{axcurrlo}). Thus, 
 according to our chiral power counting, we have three types of NLO
corrections to $A_{\mu\nu}$. 
The first is the NLO correction to the pion 
 propagator  we have evaluated in section \ref{nloprop}, coming from  
 the product of the $\Od (\pi)$ terms above. 
The second is the product of the $\Od (\pi)$ with the $\Od (\pi^3)$, 
 represented by  diagram c) in Fig.\ref{fig2}, and the third comes from
 the modification in $A_\mu$ due to the action (\ref{s4}), which amounts 
 to  prefactors $[1+f_1(t)]$ and $[1+f_2(t)]$ in $A_0$ and $A_j$ 
respectively. Then, evaluating  $\rho_{\mu\nu}$ to NLO, after 
using (\ref{nloposit})  
(we take, without loss of generality, 
both $t,t'\in C_1$ and positive )
and  (\ref{fpisp})-(\ref{fpits}),  we  finally arrive to
\dnote{Cut $\rho_L (\omega_q,t,t')$ formula and equations of the 
$f_\pi's$ in terms of ${\cal H}$} 
\ba
\left[\fpisp (t)\right]^2&=&f^2 (t)\left[1+2f_2(t)-f_1(t)\right]-2i G_0(t)
\label{fpispnlo}\\
\left[\fpite (t)\right]^2&=&f^2 (t)\left[1+f_2(t)\right]-2i G_0(t)
\label{fpitenlo}
\ea
for $t> 0$. This is the main result of this work. It provides  
 the NLO relationship between  the PDF
 and $\ft$ \footnote{$G_0 (t)$, $f_1(t)$ and $f_2(t)$  
depend implicitly on $f(t)$, through  (\ref{loprop}) and (\ref{f12}).}. 
Notice that 
 $\fpisp (t)\neq\fpite (t)$ to NLO, unlike the equilibrium case, 
due to the effect of nonequilibrium renormalisation. 
  However, note that 
$[\fpisp (t)]^2-[\fpite (t)]^2=f^2 (t)[f_2 (t)-f_1 (t)]$, which is 
 {\it finite}, since it depends only on $L_{12}$, which does not 
 renormalise. This is indeed an interesting consistency 
 check, because  the one-loop infinities appearing in $G_0(t)$ can 
 then be absorbed in  $L_{11}$, 
 rendering both $\fpisp (t)$ and $\fpite (t)$ finite.  We also remark that 
 we did not need to take   $\omega_q\rightarrow 0^+$ in 
(\ref{fpisp})-(\ref{fpits}) to arrive to (\ref{fpispnlo})-(\ref{fpitenlo}) 
(there are  still NGB to NLO).  
Note also that both $f_\pi^{s,t}$ are real 
 to this order. 
We have performed the following consistency checks on 
(\ref{fpispnlo})-(\ref{fpitenlo}): first, the 
equilibrium result (\ref{fpieq}) is recovered (for the  
 contour $C$) simply  by replacing  $G_0^{eq}=-iT^2/12$ and $f_1=f_2=0$. 
 Second, by calculating  $\gpit$ from $\rho_S$, through 
 (\ref{fpisgpi}),  we check explicitly that 
  the WI  (\ref{wifpi}) holds and, third, we have calculated $A_{\mu\nu}$
 in the $\tilde\pi$ parametrisation,   arriving  to the same result.
  \dnote{cut the comment about $v(t)$}
 
 Therefore,  (\ref{fpispnlo})-(\ref{fpitenlo}) 
  allow to express  nonequilibrium observables 
(like decay rates, masses, etc)  to one loop in ChPT, in 
 terms of the physical $\fpit$, which could be measured, for instance, 
 in nonequilibrium lepton decays $\pi\rightarrow l \nu_l$. At this stage 
 one can follow different approaches. 
  Exact 
 knowledge of  $\ft$ would require to solve  self-consistently 
 the plasma hydrodynamic  equations or,  equivalently,  
Einstein equations for the metric. Alternatively, one can  treat 
 $\ft$ as external ---so that (\ref{fpispnlo})-(\ref{fpitenlo}) provide
 the system response--- 
  and  study simple choices   consistently   with  (\ref{chipoco}) 
 \cite{agnwip}. 
In what follows, we shall take $\ft$ arbitrary and expand it 
 near $t=0^+$,  analysing  thus the short-time evolution.

 For short times, the particular form of $\ft$ is not important and 
 we can  parametrise the  nonequilibrium dynamics 
in terms of  the values of $\ft$ and 
 its derivatives at $t=0^+$.  As we discussed in 
 section \ref{model}, this approach is justified for 
times $t<t_{max}$ with 
 $t_{max}\simeq 1/\fpi (0)\simeq$ 2 fm/c (compare
 to the typical plasma time scales  5-10 fm/c \cite{cooper95}).
The general  solution of (\ref{lopropmix}) with KMS conditions at $t_i$ 
 can be  constructed  in terms of two independent 
solutions to the homogeneous equation, which have to be  continuous and 
differentiable $\forall t\in C$ so that the solution is 
 uniquely defined  \cite{sewe85}. Therefore, they have to match the 
equilibrium solution and its first time  derivative at $t=0$. 
 With these  conditions and 
expanding both  $\ft$ and the solutions near  $t=0^+$ 
 we find  to the  lowest order
\dnote{Cut details of the calculation, $g_i$'s, Wronskian.} 
\be
G_0^{11}(\omega_q,t,t)=-\frac{i}{2\omega_q}
\coth\left[\frac{\beta_i\omega_q}{2}\right]\left[1-m^2t^2 + 
\Od (m^4t^4)\right]
\label{g0st}
\ee
for $t>0$, 
with $m^2=-\ddot f(0^+)/f(0^+)$. For $m^2<0$ 
we see  the unstable modes threshold, 
 making the pion correlation function grow with time. The effect
 of those modes is not important for short times though, where the exponential
 growth of the correlator is not appreciable. 
 Observe that in (\ref{g0st}) 
 the time dependence factorises, so that the momentum dependence 
 is the same as in equilibrium and then we can  integrate it in 
DR, yielding the finite answer (\ref{nlogeq}).
 Then, from (\ref{fpispnlo})-(\ref{fpitenlo})  we get
\be
\left[\fpi^{s,t}(t)\right]^2=\left[f_R^{s,t}\right]^2
\left\{1-\frac{T_i^2}{T_c^2}
+2Ht
-\left[m^2\left(1-\frac{T_i^2}{T_c^2}\right)-H^2\right]t^2
+\Od(p^3/\Lambda_\chi^2)\right\}
\label{nlocoeff}
\ee
 for $t>0$, where  $H=\dot f(0^+)/f(0)$, with  the renormalised constants 
 \begin{eqnarray}
\left[f_R^s \right]^2&=&f^2 (0^+)+
4\left[(L_{12}-6L_{11})m^2-L_{12}H^2\right]
\nonumber\\
\left[f_R^t\right]^2&=&f^2 (0^+)+4\left[-(L_{12}+6L_{11})m^2+L_{12}H^2\right]
\label{f+ren}   
 \end{eqnarray}
 and where  the $H$ and $m$ parameters (which are
 $\Od(p)$ and play the role of the Hubble constant and the deceleration 
 parameter in the Universe expansion) also get 
 renormalised, in terms of $\dot f_{1,2}(0^+)$, $\ddot f_{1,2}(0^+)$ and 
so on, but those are subleading contributions.  Thus, for 
 short times, all the effect of the $S_4$ terms, which is $T_i$ 
 independent,  is to redefine 
$f_\pi (0^+)$, since there are no infinities coming from $G_0$ in DR. 
 We insist 
 that this is just the effect of truncating the series in $t$ and it is 
 not true in general. 
Notice that  (\ref{f+ren}) implies 
necessarily  a nonzero jump $\Delta f=f(0^+)-f$ (see our comments in 
 section \ref{model}) so that the  
 divergent part of $L_{11}$ can be absorbed in $f^2 (0^+)$ 
rendering a finite $\Delta f_R^{s,t}=\Delta f_\pi^{s,t}$. In fact, 
 that effect  is very small compared to the other contributions in 
 (\ref{nlocoeff}), and so it is the 
 difference $\fpisp (t)-\fpite(t)$,  
since  $L_{11}^r,L_{12}^r\simeq 10^{-3}$ \cite{dole91}.
  Notice also that for the particular case 
$H^2=m^2>0$,  renormalising as  $f^2(0^+)=f^2+24m^2L_{11}$ 
we get $\fpisp=\fpite$ and $\Delta_{f_\pi}=0$.

 Finally, we will estimate some  physical effects related to 
 $\fpit$. For that purpose, 
 we will ignore, for simplicity, the effect of $L_{11}$ and $L_{12}$ and, 
 based upon (\ref{fpieq}),  
 define the plasma effective temperature as
 $\fpitsq =f^2 [1-T^2(t)/T_c^2]$. Therefore, we can also define 
  a critical  time  as $T(t_c)=T_c$ and a freezing 
 time $T(t_f)=0$.
Thus, we will impose $0<T(t)<T_c$ and then, through our  short-time 
 results for  $\fpit$, determine  either $t_c$ or $t_f$, depending on 
 the initial conditions ($T(t)$ is just quadratic in time to this order). 
Notice that we are following a  similar approach as in 
equilibrium  when one  extrapolates  (\ref{fpieq}) until  $T=T_c$. 
%
%
%
%
%
%
%
Let us then  take typical  values  
$T_i, |H|, |m|\simeq 100$ MeV and retain only the 
leading order in  $x\equiv T_i^2/T_c^2$, 
consistently  with the chiral expansion. 
Then, if $H=0$, the system cools down until $t_f^2 m^2\simeq -x(1+x)$ 
 ($t_f\simeq 0.2$ fm/c)
 if  $m^2<0$, whereas for $m^2>0$ it is heated until
 $t_c^2 m^2=1$ ($t_c\simeq 2$ fm/c), independent  of $T_i$. 
 For $H>0$, there is cooling until $t_f|H|\simeq x/2$ ($t_f\simeq 0.2$ fm/c).
 Finally, for $H<0$ and $m^2>0$ there is heating until 
 $t_c|H|\simeq (1-3x/4)/2$ ($t_c\simeq 2$ fm/c), whereas if 
 $m^2<0$, there is heating until a maximum 
$t_m|H|\simeq (1+x/2)/2$ and then cooling down until 
 $t_f|H|\simeq 1+x$ ($t_f\simeq 2.3$ fm/c). 
 We observe that the effect of the unstable modes ($m^2<0$) is always 
to cool down the system and that  the freezing time  for $H<0$  is 
  much longer  than that for $H>0$. Some of these time scales are indeed
 longer than those  to which our short-time approximation 
 remains valid, but they have to be understood as 
 estimates, similarly to estimating $T_c$  at equilibrium through 
  (\ref{fpieq}), 
even though the low $T$ approach is less reliable near 
 $T\simeq T_c$.

Comparing with  \cite{cooper95}, naively  identifying 
 the LSM order parameter $v(t)\simeq f_\pi (t)$ (in proper time), 
we see a similar 
 short-time evolution, although our estimates 
  for the  time evolution duration are somewhat lower. 
 This was expected, since the initial values in  \cite{cooper95} 
 correspond to 
 $T_i\simeq$ 200 MeV and $|H|\simeq$ 400 MeV, which are
 too high for our low-energy approach. 
 An important remark 
is that in typical simulations like \cite{cooper95}, 
$v (t)$ reaches a stationary value, about which it 
 oscillates (thermalisation). 
 It is clear that we cannot predict that type of behaviour only 
 within  our short-time approach, quadratic in time, but only estimate the 
 time scales involved ---similarly as to why
ChPT cannot see the phase transition---. Therefore, in view of the 
 above estimates,  we believe that our ChPT
 model may be useful for studying the different 
 nonequilibrium observables evolution, from 
 a stage where some cooling has already taken place onwards. 
 In principle, we could 
 approach closer to $T_c$ by considering  enough orders in our ChPT, 
 although 
 in practice, beyond one-loop, some resummation method, like large $N$, 
 will need to be implemented.  
\section{Conclusions and Outlook}
\label{conc} 
We have extended   chiral lagrangians and ChPT  
 out of thermal equilibrium. 
The chiral power counting requires  
 all time derivatives to be $\Od (p)$ and  to lowest order
 our model is a 
 NLSM with $f\rightarrow \ft$. This model 
 accommodates unstable pion modes and 
 corresponds to a spatially flat RW metric in conformal time 
 with scale factor $a(t)=f(t)/f(0)$ and minimal coupling. We have exploited 
 this analogy to establish the renormalisation procedure, which allows
  to  construct the fourth order lagrangian absorbing  
 all the loop divergences, which in general will be time-dependent. 

 We have applied our  model to study 
 the  time-dependent pion decay functions, extending  the equilibrium 
 pion decay constants. In general there are two 
 independent PDF, as in   equilibrium, and  to NLO in ChPT they already 
 differ, unlike equilibrium, due to renormalisation.  
We  have obtained them  to NLO in terms 
 of the equal-time correlation function, analysing  their lowest order 
short-time coefficients and their dependence with $T_i$, and 
 discussing the  relevant time  scales  involved within the context of 
 a  RHIC plasma.

Among  the aspects of our model which are worth  to be studied  further are 
 the  long-time evolution, by choosing suitable parametrisations for $\ft$,
 including the \dnote{adiabatic?} analytic approach,  
 and  the behaviour of the two-point correlation function 
at different space points, which would allow us to investigate 
 the formation 
 of regions of unstable vacua (DCC) \cite{agnwip}.  
 Other applications and extensions, 
to be explored in the future include  photon production in the pion 
sector (by gauging the theory and including  $\pi^0$ anomalous 
decay),   the quark condensate 
 time dependence (by including the mass explicit symmetry-breaking terms), 
 the $N_f=3$ case, large $N$ resummation and  proper time evolution. 
\section*{Acknowledgments}
We are grateful to T.Evans and  R.Rivers  for countless
  and fruitful  discussions, as well as to 
 R.F.Alvarez-Estrada, A.Dobado and A.L.Maroto
 for providing useful references and comments. A.G.N wishes to thank the 
 Imperial College group for their kind hospitality during his stay there 
 and has received financial support through  a postdoctoral 
 fellowship of the Spanish Ministry of Education and 
 CICYT, Spain, project AEN97-1693. 
.












\end{document}